\documentstyle[12pt]{article}

\def\hybrid{\topmargin -20pt \oddsidemargin 0pt
\headheight 0pt \headsep 0pt
\textwidth 6.25in % A4 paper
\textheight 9.5in % A4 paper
\marginparwidth .875in
\parskip 5pt plus 1pt \jot = 1.5ex}

\hybrid

\def\baselinestretch{1.2}

\catcode`\@=11

\def\marginnote#1{}
\newcount\hour
\newcount\minute
\newtoks\amorpm
\hour=\time\divide\hour by60
\minute=\time{\multiply\hour by60 \global\advance\minute by-\hour}
\edef\standardtime{{\ifnum\hour<12 \global\amorpm={am}%
\else\global\amorpm={pm}\advance\hour by-12 \fi
\ifnum\hour=0 \hour=12 \fi
\number\hour:\ifnum\minute<10 0\fi\number\minute\the\amorpm}}
\edef\militarytime{\number\hour:\ifnum\minute<10 0\fi\number\minute}
\def\draftlabel#1{{\@bsphack\if@filesw {\let\thepage\relax
\xdef\@gtempa{\write\@auxout{\string
\newlabel{#1}{{\@currentlabel}{\thepage}}}}}\@gtempa
\if@nobreak \ifvmode\nobreak\fi\fi\fi\@esphack}
\gdef\@eqnlabel{#1}}
\def\@eqnlabel{}
\def\@vacuum{}
\def\draftmarginnote#1{\marginpar{\raggedright\scriptsize\tt#1}}

\def\draft{\oddsidemargin -.5truein
\def\@oddfoot{\sl preliminary draft \hfil
\rm\thepage\hfil\sl\today\quad\militarytime}
\let\@evenfoot\@oddfoot \overfullrule 3pt
\let\label=\draftlabel
\let\marginnote=\draftmarginnote
\def\@eqnnum{(\theequation)\rlap{\kern\marginparsep\tt\@eqnlabel}%
\global\let\@eqnlabel\@vacuum} }

\def\preprint{\twocolumn\sloppy\flushbottom\parindent 2em
\leftmargini 2em\leftmarginv .5em\leftmarginvi .5em
\oddsidemargin -.5in \evensidemargin -.5in
\columnsep .4in \footheight 0pt
\textwidth 10.in \topmargin -.4in
\headheight 12pt \topskip .4in
\textheight 6.9in \footskip 0pt
\def\@oddhead{\thepage\hfil\addtocounter{page}{1}\thepage}
\let\@evenhead\@oddhead \def\@oddfoot{} \def\@evenfoot{} }

\def\numberbysection{\@addtoreset{equation}{section}
\def\theequation{\thesection.\arabic{equation}}}

\def\underline#1{\relax\ifmmode\@@underline#1\else
$\@@underline{\hbox{#1}}$\relax\fi}

\def\titlepage{\@restonecolfalse\if@twocolumn\@restonecoltrue\onecolumn
\else \newpage \fi \thispagestyle{empty}\c@page\z@
\def\thefootnote{\fnsymbol{footnote}} }

\def\endtitlepage{\if@restonecol\twocolumn \else \newpage \fi
\def\thefootnote{\arabic{footnote}}
\setcounter{footnote}{0}} %\c@footnote\z@ }

\catcode`@=12
\relax

\def\figcap{\section*{Figure Captions\markboth
{FIGURECAPTIONS}{FIGURECAPTIONS}}\list
{Figure \arabic{enumi}:\hfill}{\settowidth\labelwidth{Figure
999:}
\leftmargin\labelwidth
\advance\leftmargin\labelsep\usecounter{enumi}}}
 \relax
\def\tablecap{\section*{Table Captions\markboth
{TABLECAPTIONS}{TABLECAPTIONS}}\list
{Table \arabic{enumi}:\hfill}{\settowidth\labelwidth{Table
999:}
\leftmargin\labelwidth
\advance\leftmargin\labelsep\usecounter{enumi}}}
 \relax
\def\reflist{\section*{References\markboth
{REFLIST}{REFLIST}}\list
{[\arabic{enumi}]\hfill}{\settowidth\labelwidth{[999]}
\leftmargin\labelwidth
\advance\leftmargin\labelsep\usecounter{enumi}}}
 \relax
\makeatletter
\newcounter{pubctr}
\def\publist{\@ifnextchar[{\@publist}{\@@publist}}
\def\@publist[#1]{\list
{[\arabic{pubctr}]\hfill}{\settowidth\labelwidth{[999]}
\leftmargin\labelwidth
\advance\leftmargin\labelsep
\@nmbrlisttrue\def\@listctr{pubctr}
\setcounter{pubctr}{#1}\addtocounter{pubctr}{-1}}}
\def\@@publist{\list
{[\arabic{pubctr}]\hfill}{\settowidth\labelwidth{[999]}
\leftmargin\labelwidth
\advance\leftmargin\labelsep
\@nmbrlisttrue\def\@listctr{pubctr}}}
 \relax
\makeatother
\newskip\humongous \humongous=0pt plus 1000pt minus 1000pt

\newif\ifdtup

\relax

\def\be{\begin{equation}}
\def\ee{\end{equation}}
\def\ba{\begin{eqnarray}}
\def\ea{\end{eqnarray}}

\def\del{\partial}

\def\a{\alpha}

\def\b{\beta}

\def\D{\Delta}

\def\th{\theta}

\def\m{\mu}
\def\n{\nu}

\def\no{\noindent}

\def\qq{\qquad}

\def\IR{\relax{\rm I\kern-.18em R}}

\def \ha {{1\over 2}}

\def \ov {\over}

\def\IR{\relax{\rm I\kern-.18em R}}
\def\inv{^{\raise.15ex\hbox{${\scriptscriptstyle -}$}\kern-.05em 1}}

\def\tL{{\tilde L}}

\begin{document}
%\draft

\renewcommand{\theequation}{\arabic{equation}}

\newcommand{\beq}{\begin{equation}}
\newcommand{\eeq}[1]{\label{#1}\end{equation}}
\newcommand{\ber}{\begin{eqnarray}}
\newcommand{\eer}[1]{\label{#1}\end{eqnarray}}
\newcommand{\eqn}[1]{(\ref{#1})}
\begin{titlepage}
\begin{center}

%Phys. Lett. {\bf B432} (1998) 365
\hfill CERN-TH/98-409\\
\hfill hep--th/9901056\\

\vskip .8in

{\large \bf Rotating D3-branes and QCD in three dimensions}

\vskip 0.4in

{ Jorge G. Russo${}^1$}\hskip .25 cm and\hskip .3cm
{ Konstadinos Sfetsos${}^2$}
\vskip 0.1in
{\em ${}^1\!$Departamento de F\' \i sica, Universidad de Buenos Aires,\\
Ciudad Universitaria, 1428 Buenos Aires, Argentina\\
{\tt russo@df.uba.ar }}\\
\vskip 0.2in
{\em ${}^2\!$Theory Division, CERN\\
CH-1211 Geneva 23, Switzerland\\
{\tt sfetsos@mail.cern.ch}}
\vskip .2in

\end{center}

\vskip .4in

\centerline{\bf Abstract }
\no
We investigate the rotating D3-brane solution with maximum number of angular 
momentum parameters. After determining the angular velocities,
Hawking temperature, ADM mass and entropy,
we use this geometry to construct general three-parameter
models of non-supersymmetric pure
$SU(N)$ Yang--Mills theories in 2+1 dimensions.
We calculate glueball masses in the WKB approximation
and obtain closed analytic expressions
for generic values of the parameters.
We  also determine the masses of Kaluza--Klein states associated with
internal parts of the ten-dimensional metric
and investigate
the parameter region where some of these states are decoupled.
To leading order in $1/\lambda $ and $1/N$ (where $\lambda $ is the
't Hooft coupling) we find a global $U(1)^3$ symmetry and
states with masses comparable
to glueball masses, which have no counterpart in the more familiar
(finite $\lambda , N$) Yang--Mills theories.

\vskip 0,2cm
\no

%\phantom{\cite{Wittads,GO98,COOT98,russo,haoz,csaki}}

\vskip 2.5 cm
\noindent
CERN-TH/98-409\\
December 1998\\
\end{titlepage}
\vfill
\eject

\def\baselinestretch{1.2}
\baselineskip 16 pt
\noindent

\def\tT{{\tilde T}}
\def\tg{{\tilde g}}
\def\tL{{\tilde L}}

%%%%%%%%%%%%%%%%%%%
\section{Introduction}

The extremal D3-brane provides one of the simplest illustrations
of the recently found dualities between gauge
theories and string theory on geometries which asymptotically
approach the Anti-de Sitter space-time \cite{mal}.
Because the  string models on these geometries are not
understood to date,
much of the attention has been devoted to those models
where the curvature of the geometry
is small everywhere, so that the system can be studied by
using supergravity.
This is the case for black brane geometries,
which for large charges have
small curvatures everywhere outside the horizon.
The most general geometry with a regular horizon, which has a
D3-brane charge, is given by the non-extremal D3-brane with
three angular momenta.
It is of obvious interest to investigate the spectra
of this general (five-parameter) supergravity model based on
the D3-brane. In this paper we determine the spectrum of
the corresponding Laplace operator in the WKB approximation.

Models of QCD in $2+1$ dimensions can be constructed from the
non-extremal D3-brane geometry by compactifying the euclidean
time direction, which plays the role of an internal angle
[2--7].
In the case of zero angular momentum,
there is a single mass scale in the spectrum (given by the
Hawking temperature); Kaluza--Klein
states associated with the compact
Euclidean time have the same mass scale as
the states with vanishing Kaluza--Klein charge (e.g. glueballs)
and the dimensional reduction is not justified.
This problem can be overcome by starting with a rotating D3-brane
with large angular-momentum, in which case the radius of the
circle shrinks to zero and the Kaluza--Klein particles decouple
\cite{russo,csaki}.
Another problem in making contact with ordinary Yang--Mills theory
was pointed out in \cite{ORT}.
While the static D3-brane has an $SO(6)$ isometry
group associated with  the five-sphere part of the geometry,
in pure non-supersymmetric Yang--Mills theory there is no
counterpart of this $SO(6)$ global symmetry.
Introducing angular momentum
 breaks the $SO(6)$ isometry group to smaller subgroups.
When the maximum number of angular momentum components are turned on,
the only remaining global symmetry is the Abelian (Cartan) subgroup
$U(1)\times U(1)\times U(1)$ of $SO(6)$.
Although this is in a sense closer to QCD than
a model based on the static D3-brane
(which implies a large $SO(6)$ global symmetry),
there is no Abelian $U(1)^3$ global symmetry in
weakly coupled non-supersymmetric $SU(N)$ Yang--Mills theory
without matter.
A natural question is whether the Kaluza--Klein particles, which
are charged with
respect to the $U(1)^3$ Cartan group, could have large masses
(with respect to the glueball masses) in some region of the
three-dimensional parameter
space $a_1,a_2,a_3$.\footnote{There are  two extra parameters
associated with charge and mass of the D3-brane:
the mass fixes the scale and can be set to 1;
the charge is related to the 't Hooft coupling $\lambda $ and
in the supergravity approximation does not affect the mass spectra.}
In Sect.~4.3 it will be shown that in the supergravity approximation
these Kaluza--Klein particles cannot
decouple in any region of the parameter space $a_1,a_2,a_3$.
It should be noted that the WKB approximation is sufficient
to establish that certain states do not decouple, since
one can consider in particular states with sufficiently high radial quantum
numbers so that
the WKB approximation is close to the exact result.
It is nevertheless reasonable to expect
that these exotic Kaluza--Klein particles may decouple once all $1/\lambda $
effects are incorporated, since the weakly coupled
non-supersymmetric Yang--Mills  theory that should govern the low-energy
dynamics of the compactified
D3-brane does not have these particles in the physical spectrum.
A further discussion on this is given in Sect.~5.

% A similar discussion for QCD in $3+1$ dimensions
% is given.

\section{The rotating D3-brane background}
%%%%%%%%%%%%%%%%%%%%%%%%%%%%%%%%%%

The D3-brane has an internal $SO(6)$ rotational isometry,
which allows three independent angular parameters, $l_1, l_2,l_3$.
The metric for the case $l_2=l_3=0$ was given in \cite{russo}.
The  general metric with parameters $l_1, l_2,l_3$
was recently given in an appendix in
\cite{KLT} and was obtained by duality transformations of the black
hole solutions of \cite{cvetic} (there are two typographical errors
in the expression of \cite{KLT} that are corrected below).
The rotating D3-brane metric is given by
\ba
ds^2_{\rm IIB} &=&  f^{-1/2}_0 \big(  -h_0 dx_0^2
+dx_1^2+dx_2^2+dx_3^2\big)
+f^{1/2}_0 \bigg[{\Delta\ dr^2\over \prod_{i=1}^3 (1+{l_i^2\over
r^2} ) -{2m\over
r^4} }
\nonumber\\
&+& r^2
\bigg(\Delta_1 d\theta^2 +\Delta_2 \cos^2\theta d\psi^2 -
2 {l_2^2-l_3^2\over r^2}\cos\theta\sin\theta\cos\psi\sin\psi d\theta d\psi
\nonumber\\
&+& (1+{l_1^2\over r^2})\sin^2\theta d\varphi_1^2 +
(1+{l_2^2\over r^2}) \cos^2\theta \sin^2\psi d\varphi_2^2 +
(1+{l_3^2\over r^2}) \cos^2\theta\cos^2\psi d\varphi_3^2
\nonumber\\
&+& {2m\over r^6\Delta f_0}\big( l_1\sin^2\theta d\varphi_1+
l_2 \cos^2\theta \sin^2\psi d\varphi_2
+ l_3 \cos^2\theta\cos^2\psi d\varphi_3 \big)^2
 \bigg)
\nonumber\\
&-& {4 m \cosh\a \over r^4\Delta f_0}dx_0 \big( l_1\sin^2\theta d\varphi_1
+l_2\cos^2\theta\sin^2\psi d\varphi_2+ l_3\cos^2\theta\cos^2\psi
d\varphi_3 \big) \bigg] \ ,
\label{aiib}
\ea
where
\ba
\Delta &=& 1 +{l_1^2\over r^2} \cos^2\theta +{l_2^2\over r^2}
(\sin^2\theta\sin^2\psi +\cos^2\psi )
+{l_3^2\over r^2}(\sin^2\theta\cos^2\psi +\sin^2\psi )
\nonumber \\
&+& {l_2^2l_3^2\over r^4}\sin^2\theta +{l_1^2 l_3^2\over r^4}
\cos^2\theta\sin^2\psi +{l_1^2l_2^2\over r^4}\cos^2\theta\cos^2\psi\ ,
\nonumber \\
\Delta_1 &=& 1+{l_1^2\over r^2}\cos^2\theta +
{l_2^2\over r^2}\sin^2\theta\sin^2\psi +
{l_3^2\over r^2}\sin^2\theta\cos^2\psi\ ,
\nonumber \\
\Delta_2 &=& 1+{l_2^2\over r^2}\cos^2\psi +{l_3^2\over r^2}\sin^2\psi\ ,
\nonumber\\
h_0 & = & 1-{2m \ov r^4 \Delta}\ ,  \  \  \  \  f_0=1+{2m \sinh^2\a\over
r^4\Delta}\ .
\label{ad12}
\ea
The dilaton field $\Phi $ is constant, $e^\Phi =g_s$. The angular part
differs from
eq.~(69) of \cite{KLT} in
the components $G_{\theta \theta}$ and $G_{\psi\psi}$.
Some interesting thermodynamical aspects of rotating D3-branes
have recently been investigated in refs.~\cite{gubser,KLT,caso}
in the context  of gauge/string-theory correspondence, which indicate
that there may be a phase transition in ${\cal N}=4$ super-Yang--Mills theory
at finite temperature \cite{gubser,caso}.
At zero temperature the supergravity solution corresponds to the continuum
limit \cite{KLT,sfetsos} 
of a multicenter static D3-brane solution and describes
a Higgs phase of ${\cal N}=4$ super-Yang--Mills theory with a vacuum having a
$Z_N$-type symmetry \cite{sfetsos}.
It would be interesting to extend these discussions to the general
case (\ref{aiib}) of maximum number  of angular momenta.
The interpretation is somewhat different in the present case, where
(\ref{aiib}) will be used to construct a static space-time, with the
Euclidean time
parametrizing an internal circle.

The parameters $m$ and $\a $ are related to the D3-brane 
charge $N$ by
%\be
%2m\cosh\a \sinh\a =4\pi g_s N{\alpha ' }^2\ .
%\label{mmn}
%\ee
\be
\sinh^2\a = \sqrt{(2 \pi g_s N \a'^2/m)^2 +1/4} -1/2 \ .
\label{mm1}
\ee
For completeness we also include the 4-form gauge field \cite{KLT}
\ba
C^{(4)} & =& {1-f_0^{-1}\ov \sinh \a} \ dx_1 \wedge dx_2 \wedge dx_3 \wedge
(\cosh \a\ dt - l_1 \sin^2\th\ d\varphi_1 \\
\nonumber
&& -l_2 \cos^2\th \sin^2\psi\ d\varphi_1
-l_3 \cos^2\th \cos^2\psi\ d\varphi_1 )\ .
\label{c4fo}
\ea

The location of the horizon $r=r_H$ is given by the largest real root of
\be
\prod _{i=1}^3 (r^2+l_i^2) -2m r^2=0\ ,
\label{hori}
\ee
which is a cubic equation for $r^2$.
The angular velocities $\Omega_1 ,\ \Omega_2,\ \Omega_3 $ associated with
motion in
$\varphi_1 ,\ \varphi_2,\ \varphi_3$ can be
determined by requiring that the vector
$\eta ={\partial\over\partial x_0}+\Omega_i {\partial\over\partial
\varphi_i}$
be null at the horizon. They are independent of the angles,
so that one can compute them by evaluating $\eta ^2$ at different values of
$\theta $ and
$\psi $. At
$\theta=0,\psi=0$, $\eta ^2 $ is independent of $\Omega_{1,2}$, so that
setting $\eta^2 =0 $ determines $\Omega_3 $. Similarly, by evaluating
$\eta^2$
at $\theta=0, \psi=\pi/2 $, and using the value
of $\Omega_3 $ already obtained, one finds $\Omega_2 $,
whereas by evaluating it at $\theta=\pi/2 ,\ \psi=\pi/2 $ one obtains
$\Omega_1 $. The result can be written compactly as
\be
\Omega_i={l_i\over  \cosh\a \big( r_H^2+l_i^2 \big) } \ ,\ \ \ \ \ i=1,2,3\
.
\ee
%%%%We used \be \eta ^2= \ee
The Hawking temperature is obtained from $\eta ^2 $ by the formula
\be
T_H^2=  \lim_{r\to r_H} {1\over 16\pi^2 (-\eta ^2 )}
\nabla_\mu \eta ^2 \nabla^\mu \eta ^2\ .
% = -\lim_{r\to r_H} {1\over 16\pi^2 \eta ^2 } (\partial_r \eta ^2 )^2
\ee
Being constant, it can be computed at
$\theta =\pi/2 ,\psi=0 $.
After a somewhat long but straightforward calculation we find
\ba
T_H & = & {r_H\over 4\pi m\cosh\a }
\Bigg(
2r_H^2+l_1^2+l_2^2+l_3^2 - {l_1^2l_2^2l_3^2\over r_H^4}\Bigg)
\nonumber \\
& =& {1\over 4 \pi r_H m\cosh\a } \big(r_H^2-r_1^2\big)
\big(r_H^2-r_2^2\big)        \ ,
\label{hawr}
\ea
where $r_H^2, r_1^2,r_2^2$ are the
three roots of eq.~(\ref{hori}).
The second line in \eqn{hawr} can be proved
by multiplying \eqn{hori} by $(r^2-r_H^2)^{-1}$ and
taking the limit
$r^2\to r^2_H$.
The  mass, entropy and angular momenta are given by
\ba
M_{\rm ADM}&=&{V_3 V(\Omega_5) \over 4 \pi G_N} 
2m \big( {5\over 4}+\sinh^2\a \big)\ ,\ \ \ \ \ V(\Omega_5)=\pi^3\ ,
\\
S&=&{V_3 V(\Omega_5) \over 4G_N} 2mr_H \cosh\a\ ,
\\
J_i&=&{V_3 V(\Omega_5) \over 4\pi G_N} m l_i \cosh\a\ ,\ \ \ \ i=1,2,3\ ,
\label{masa}
\ea
where $V_3$ is the volume of the 3-brane.

\section{WKB method}

Glueball masses in the models of \cite{Wittads,russo}
have been calculated in the WKB approximation in \cite{minahan}.
By extending the approach of \cite{minahan},
we develop in this section a  simple formalism that will be
useful to calculate the different mass spectra (including Kaluza--Klein
 modes) in the present case of three angular momenta.

We are interested in differential equations of the form
\be
\del_u \left(f(u) \del_u \phi\right) + \left(M^2 h(u)
+ p(u)\right) \phi = 0 \ ,
\label{diiff}
\ee
where $M$ represents a mass parameter, and
$f(u)$, $h(u)$ and $p(u)$ are three arbitrary functions
that are independent of $M$ and have the following behavior.
%depending on the detail
%structure of the metric and $M$ is the glueball mass.
There is a point $u_H$, where
\be
f\approx f_1 (u-u_H)^{s_1}\ ,\quad h\approx h_1 (u-u_H)^{s_2}\ ,
\quad p\approx p_1 (u-u_H)^{s_3}\ , \qq {\rm as}\quad
u\to u_H\ ,
\label{lii0}
\ee
for some constants $s_1$, $s_2$, $s_3$, $f_1$, $h_1$ and $p_1$.
Similarly, we assume that
\be
f\approx f_2 u^{r_1}\ ,\quad h\approx h_2 u^{r_2}\ ,
\quad p\approx p_2 u^{r_3}\ ,\qq {\rm as}\quad u\to \infty\ ,
\label{lii1}
\ee
for some other constants $r_1$, $r_2$, $r_3$, $f_2$, $h_2$
and $p_2$.
For large masses $M$, one
may apply WKB methods to obtain the approximate spectrum and expressions for
$\phi$. In order to apply standard formulae from the WKB-approximation
theory
we cast (\ref{diiff}) into the form of a Schr\" odinger
equation
\be
\del_y^2\psi + V(y)\psi = 0\ .
\label{schrr}
\ee
The
necessary transformation that brings (\ref{diiff}) into the form
(\ref{schrr}) is
\be
e^y = u-u_H\ ,\qq \phi = e^{y\ov 2} f^{-\ha} \psi\ ,
\label{trass}
\ee
with the potential given by
\ba
&& V(y) = M^2 {h_0\ov f_0} -\ha {f_0^{\prime\prime}\ov f_0} +
{1\ov 4} {{f_0^\prime}^2\ov f_0^2} + {p_0\ov f_0}\ ,
\nonumber \\
&& f_0\equiv e^{-y} f \ , \quad h_0 \equiv e^y h\ ,\quad
p_0 = e^y p \ .
\label{trass1}
\ea
\no
{}From the asymptotic expressions (\ref{lii0}) and (\ref{lii1}) one
finds that

\be
V(y)\approx {h_1\ov f_1} M^2 e^{(s_2-s_1+2) y} + {p_1\ov f_1}
e^{(s_3-s_1+2)y } - {1\ov 4}(s_1-1)^2\ ,\qq {\rm for}
\quad y\ll 0 \ ,
\label{ass1}
\ee
and
\be
V(y)\approx {h_2\ov f_2}
M^2 e^{-(r_1 -r_2-2) y} + {p_2\ov f_2} e^{-(r_1-r_3-2)y}
- {1\ov 4}(r_1-1)^2\ ,\qq {\rm for} \quad y\gg 0 \ .
\label{asss}
\ee

\no
Consistency requires that $s_2-s_1+2$ and $r_1 -r_2-2$ 
are strictly positive numbers, whereas $s_3-s_1+2$ and
$r_1-r_3-2$ can be either positive or zero (see also below).
{}From these expressions we see that there are two turning points $y_1$ and
$y_2$, given by solving $V(y_1)=0$ in (\ref{ass1}) and $V(y_2)=0$
in (\ref{asss}). To the order of approximation that we
will be interested for the computation of $M$, we have

\be
y_1 = -{2\ov \a_1} \ln\Bigg({2 \sqrt{h_1\ov f_1} \ov \a_2} M \Bigg)\ ,
\qq
y_2 = {2\ov \b_1} \ln\Bigg({2 \sqrt{h_2\ov f_2} \ov \b_2} M \Bigg)\ ,
\label{tuurn1}
\ee
where
\be
\a_1= s_2-s_1 +2 \ , \qq \b_1= r_1-r_2-2 \ ,
\label{a1b1}
\ee
and
\ba
&&\a_2 = |s_1-1| \quad {\rm or} \quad \a_2=\sqrt{(s_1-1)^2 -4 {p_1\ov f_1} }
\
\
({\rm if}\ s_3-s_1+2=0)\ ,
\nonumber \\
&& \b_2 = |r_1-1| \quad {\rm or} \quad \b_2=\sqrt{(r_1-1)^2 -4 {p_2\ov f_2}}
\ \
({\rm if}\ r_1-r_3-2=0)\ .
\label{a2b2}
\ea

\no
Then the mass spectrum is computed using the standard WKB formula
\ba
\left(m-\ha\right) \pi = \int_{y_1}^{y_2} dy \sqrt{V(y)} \ ,\qq m \geq 1\ .
\label{wkbm}
\ea
One may expand the right-hand side as a power series in ${1\ov M}$. The
leading term is ${\cal O}(M)$ and
is obtained by keeping, in the expression for $V(y)$ in (\ref{trass1}),
only the first term, and integrating $y$
from $-\infty$ to $+\infty$. One obtains
\be
(m-\ha) \pi =M \int_{-\infty}^{\infty}dy \sqrt{{h_0\ov f_0}}
= M \int_{u_H}^{\infty} du \sqrt{{h\ov f}} \equiv M \xi\ ,
\label{lee}
\ee

\no
where the last equality defines the constant $\xi$ with scale 
dimension of length.
The first correction of order
${\cal O}(M^0)$ has a contribution from the term
\be
- \sqrt{h_1\ov f_1} M \int_{-\infty}^{y_1}dy e^{\ha \a_1 y}
- \sqrt{h_2\ov f_2} M \int_{y_2}^{\infty}dy e^{-\ha \b_1 y}
= -{a_2\ov a_1} - {b_2\ov b_1}\ ,
\label{coorr}
\ee

\no
representing the correction that accounts for the extension of
the limits of integration from
$y_{1,2}$ to $\mp \infty$.
There is another subleading contribution from around the turning
points
\ba
&&
\int_{y_1}^{\infty} dy \Bigg( \sqrt{{h_1\ov f_1}M^2 e^{-\b_1 y} -{\b_2^2\ov
4}}
-\sqrt{h_1\ov f_1} M e^{-\ha \b_1 y} \Bigg)
\nonumber\\
&+& \int_{-\infty}^{y_2} dy \Bigg( \sqrt{{h_2\ov f_2}
M^2 e^{\a_1 y} -{\a_2^2 \ov 4} } -\sqrt{h_2\ov f_2} M e^{\ha \a_1 y} \Bigg)
= \ha (2-\pi) \Bigg( {\a_2\ov \a_1}
+ {\b_2\ov \b_1} \Bigg)\ .
\label{coorr1}
\ea

\no
Combining everything we find that
%\be
%M^2 \xi^2 = \pi^2 m \left(m -1+ {|s_1-1|\ov s_2-s_1+2}
%+ {|r_1-1|\ov r_1-r_2-2} \right)\ +\ {\cal O}(m^0) \ ,\qq m\geq 1\ .
%\label{mxi}
%\ee
\be
M^2  = {\pi^2\ov \xi^2}\ m\left(m -1+ {\a_2\ov \a_1}
+ {\b_2\ov \b_1} \right)\ +\ {\cal O}(m^0) \ ,\qq m\geq 1\ .
\label{mxi}
\ee

\no
The validity of the WKB approximation requires 
that $\a_2/\a_1 + \b_2/\b_1\ll m$
and that the  turning
points, as computed using \eqn{tuurn1}, be large in magnitude.
This gives the conditions
\be
m \left(m -1+ {\a_2\ov \a_1}
+ {\b_2\ov \b_1} \right)\ \gg \ {f_1 \a_2^2  \ov h_1} \xi^2 u_H^{-\a_1}
\quad
{\rm and} \quad {f_2 \b_2^2  \ov h_2} \xi^2 u_H^{\b_1}\ .
\label{jsdfg}
\ee

%
%Let us now consider the special case
% $s_2-s_1+2=0$ or (and) $r_1-r_2-2=0$
%(corresponding to some specific choice $f,h$ in
%(\ref{diiff})~).
%In this case eqs.~(\ref{asss})-(\ref{mxi}) do not apply, and
%one must take into account subleading terms in (\ref{lii0})
% and (\ref{lii1}) in order to derive the asymptotic
%formulae (\ref{asss}).
%Then eqs.~(\ref{asss}) and (\ref{tuurn}) are replaced by
%\ba
%V(y) &\approx & M^2 e^{a_1 y} - {1\ov 4}a_2^2\ ,\qq {\rm for} \quad
%y\ll 0 \ ,
%\nonumber \\
%V(y) &\approx & M^2 e^{-b_1 y} - {1\ov 4}b_2^2\ ,\qq {\rm for} \quad
%y\gg 0 \ ,
%\label{asss1}
%\ea
%and
%\ba
%y_1 & = & -{2\ov a_1} \ln\left({2 M\ov |a_2|}\right)\ ,
%\nonumber\\
%y_2 & = & + {2\ov b_1} \ln\left({2 M\ov |b_2|}\right)\ .
%\label{tuurn1}
%\ea
%for some constants $a_1,a_2,b_1,b_2$, where for consistency $a_1>0$ and
%$b_1>0$.
%In this case eq.~(\ref{mxi}) is replaced by
%\be
%M^2 \xi^2 = \pi^2 m \left(m -1+ {|a_2|\ov a_1}
%+ {|b_2|\ov b_1} \right)\ +\ {\cal O}(m^0) \ ,\qq m\geq 1\ .
%\label{mxi1}
%\ee

%%%%%%%%%%%%%%%%%%%%%%%%%%%%%%%%%%
\section{Glueball masses in ${\rm QCD}_3$}
%%%%%%%%%%%%%%%%%%%%%%%%%%%%%%%%%%

The three-parameter ${\rm QCD}_3$ model is obtained from the rotating 
D3-brane metric \eqn{aiib} in the following way.
We first go to Euclidean space by letting
$t\to -i \tau$ and $l_i\to i l_i$,
$i=1,2,3$.
Then we take the ``field-theory" limit \cite{mal,russo}
\be
U={r\ov \a'} \ ,\qq U_0^4={2m\ov \a'^4} \ , \qq
a_i = {l_i\ov \a'}\ ,\qq \a' \to 0\ .
\label{liim}
\ee

\no
The rotating D3-brane metric in this limit takes the form
\ba
ds^2_{\rm IIB} &=&\a ' \Delta^{1/2}
\bigg[ 
{U^2\over R^2}\big[   \big( 1-{U_0^4\ov U^4 \Delta}\big) d\tau^2
+dx_1^2+dx_2^2+dx_3^2\big]
+{R^2 dU^2\over U^2 \big[ \prod_{i=1}^3
(1-{a_i^2\over U^2})- {U_0^4\over U^4} \big] }
\nonumber\\
&+& {R^2\over \Delta }
\bigg(\Delta_1 d\theta^2 +\Delta_2 \cos^2\theta d\psi^2 +
2 {a_2^2-a_3^2\over U^2}\cos\theta\sin\theta\cos\psi\sin\psi d\theta d\psi
\nonumber\\
&+& (1-{a_1^2\over U^2})\sin^2\theta d\varphi_1^2 +
(1-{a_2^2\over U^2})
\cos^2\theta \sin^2\psi d\varphi_2^2 +
(1-{a_3^2\over U^2})\cos^2\theta\cos^2\psi d\varphi_3^2 \bigg)
\nonumber\\
&-& {
2U_0^2\over U^2\Delta }d\tau \big( a_1\sin^2\theta d\varphi_1
+a_2\cos^2\theta\sin^2\psi d\varphi_2+ a_3\cos^2\theta\cos^2\psi
d\varphi_3 \big) \bigg] \ ,
\label{dsiib}
\ea

\no
where $R^2=\sqrt{4\pi g_s N}={\rm fixed}$, and
\ba
\Delta &=& 1 -{a_1^2\over U^2} \cos^2\theta -{a_2^2\over U^2}
(\sin^2\theta\sin^2\psi +\cos^2\psi )
- {a_3^2\over U^2}(\sin^2\theta\cos^2\psi +\sin^2\psi )
\nonumber \\
&+& {a_2^2a_3^2\over U^4}\sin^2\theta +{a_1^2 a_3^2\over U^4}
\cos^2\theta\sin^2\psi +{a_1^2a_2^2\over U^4}\cos^2\theta\cos^2\psi\ ,
\nonumber \\
\Delta_1
 &=& 1-{a_1^2\over U^2}\cos^2\theta -
{a_2^2\over U^2}\sin^2\theta\sin^2\psi -
{a_3^2\over U^2}\sin^2\theta\cos^2\psi\ ,
\nonumber \\
\Delta_2 &=& 1-{a_2^2\over U^2}\cos^2\psi -{a_3^2\over U^2}\sin^2\psi\ .
\label{d12}
\ea
For large $U$ the space becomes asymptotically $AdS_5 \times S^5$, each
factor having radius
$R$ (in string units).
In the  limit \eqn{liim} the Hawking temperature is given by
\ba
T_H &=& {U_H\over 2\pi R^2 U_0^2} \Bigg(2U_H^2 - a_1^2 -a_2^2 -a_3^2 +
{a_1^2a_2^2a_3^2\over U_H^4} \Bigg)\
\nonumber\\
&=& {1\over 2\pi R^2 U_0^2U_H} \big(U_H^2 -U_1^2\big)
\big(U_H^2 - U_2^2 \big) \ ,
\label{haw}
\ea
$U_H^2$ being the largest real root of $f(U)=0$:
\be
f= \prod_{i=1}^3 (U^2-a_i^2) -U_0^4 U^2 = (U^2-U_H^2)(U^2-U_1^2)(U^2-U_2^2)
\ ,
\label{fff3}
\ee
and $U_1^2, U_2^2 $ the two other roots.
The Yang--Mills coupling  of the ($2+1$)-dimensional field theory is given by
\be
g_{\rm YM_3}^2=g^2_{\rm YM_4} T_H\ ,\ \ \ \ \  \  g^2_{\rm YM_4}=2\pi\ g_s\
{}.
\label{coup}
\ee
The 't Hooft coupling $\lambda $ is defined by
$\lambda\equiv g_{\rm YM_3}^2N/(2\pi )=T_H R^4/(4\pi) $.
%To apply the supergravity approximation the curvature must be small
%everywhere,
%in particular, at infinity. This implies $R\gg 1$.

In what follows, we will make use of the following formulas
\be
\sqrt{G}= R^2 \Delta ^{1/2} U^3 \cos^3\theta\sin\theta \sin\psi \cos\psi\ ,
\label{deet}
\ee
and
\ba
&&
G^{\psi\psi}=
{\D_1\ov R^2 \cos^2 \th \D^{1/2}}\ ,\quad G^{\th\psi}= -{a_2^2-a_3^2\ov
R^2 \D^{1/2} U^2}\ \tan\th \cos\psi \sin\psi \ ,
\nonumber\\
&& G^{\th\th}= {\D_2\ov R^2 \D^{1/2}}\ ,
\quad G^{\tau\tau}= {R^2\ov U^2 \D^{1/2}}\ {\prod_{i=1}^3 (1-{a_i^2\ov
U^2})\ov
\prod_{i=1}^3 (1-{a_i^2\ov U^2}) - {U_0^4\ov U^4}}\ .
\label{giin}
\ea

%%%%%%%%%%%%%%%%%%%%%%%%%%%%%%%%%%%%%%%
\subsection{$0^{++}$ glueballs}
%%%%%%%%%%%%%%%%%%%%%%%%%%%%%%%%%%%%%%%

Masses for $0^{++}$ glueballs are determined
from the differential equation
\be
{1\ov \sqrt{G}} \del_\m e^{-2 \Phi} \sqrt{G} G^{\m\n} \del_\n \Psi = 0 \ .
\label{dill}
\ee
We look for solutions of the form
$\Psi = \phi(u) e^{ik\cdot x}$ and identify the glueball mass square with
$M^2=-k^2$ \cite{Wittads}. After
changing variable as
$u= U^2$ (and $u_0=U_0^2$ etc.), we find an equation of the form
(\ref{diiff})
with
\be
f= \prod_{i=1}^3 (u-a_i^2) -u_0^2 u\ ,\qq h={R^4\ov 4}\ ,\qq p=0\ .
\label{fh3}
\ee
The constant $u_H$ is found by solving the cubic equation $f=0$, which can be
written as
\def\lqq{c}
\ba
&& u_H^3-\vec a^2 u_H^2 +(\vec b^2 -u_0^2)u_H - \lqq = 0 \ ,
\nonumber \\
&& \lqq= a_1^2 a_2^2 a_3^2 \ ,\qq \vec a=(a_1,a_2,a_3)\ ,\qq
\vec b=(a_2 a_3,a_1 a_3,a_1 a_2)\ .
\label{eqfl}
\ea
For the various constants necessary for the application of the WKB
method of Sect.~3, we find
\ba
&& s_1=1\ ,\quad s_2=0 \ ,\quad r_1=3 \ ,\quad r_2=0\ ,
\nonumber \\
&&\a_1= 1\ ,\quad \a_2= 0 \ , \quad \b_1= 1\ ,\quad \b_2= 2\ .
\label{smwof}
\ea
Therefore \eqn{mxi} gives
\ba
&& M^2 = {\pi^2\ov \xi^2 }\ m(m+1) \ + \ {\cal O}(m^0)\ , \qq m\ge 1\ ,
\nonumber\\
&& \xi = {R^2\ov 2} \int_{u_H}^\infty du \left(u^3-\vec a^2 u^2
+(\vec b^2-u_0^2) u - \lqq\ \right)^{-1/2}\ .
\label{qcd3xi}
\ea
The integral in \eqn{qcd3xi} can be performed explicitly.
Let $u_H, u_1$ and $u_2$ be the roots of $f=0$.
One obtains
\be
\xi = {R^2 K(k)\ov [(u_H-u_1)(u_H-u_2)]^{1/4}}\ ,
\qq k= {1\ov \sqrt{2}}
\left(1- {2u_H - u_1-u_2\ov 2 [(u_H - u_1)(u_H-u_2)]^{1/2}
}\right)^{1/2}\ ,
\label{ashg}
\ee
where $K(k)$ is the complete elliptic integral of the first kind,
i.e. $K(k)=\int_0^{\pi/2} {d\th \ov \sqrt{1-k^2 \sin^2\th}}$.
We may distinguish two cases, according to whether
$u_1$ and $u_2$ are real or  complex.
In the former case the above result can also be written as
($u_H>u_1>u_2$)
\be
\xi = {R^2 K(k_0)\ov \sqrt{u_H-u_2}}\ ,\qq
k_0 = \sqrt{u_1-u_2\ov u_H - u_2}\ .
\label{ashhh}
\ee
The validity of the WKB approximation
requires that the conditions \eqn{jsdfg}
be satisfied. These imply
\be
m(m+1) \gg \left(1\!-\!{u_1\ov u_H}\right)^{-1/2}\!
\left(1\!-\!{u_2\ov u_H}\right)^{-1/2}\! K^2(k)\ .
\label{jhdg}
\ee
%The function $(1-k^2)K^2(k)$ is bounded between
%$\pi^2/4 $ and 0.
In the case of three real roots $u_H> u_1 > u_2$,
and $u_1$ not too close to $u_H$,
$k$ is always less than 1, so that $K(k)$ is of order 1,
 and (\ref{jhdg}) implies that
the WKB approximation can be applied.
It breaks down when  $u_H\simeq u_1$.
For the case of 1 real and 2 complex roots, this happens if
${\rm Im}\ u_1 \simeq 0$ with $u_H \simeq {\rm Re} \ u_1$
(or with $u_H<{\rm Re}\ u_1$, in which case $k\simeq 1$).

If   $\lqq = a_1a_2a_3=0$ (say, $a_3=0$),
there are then three real roots --~one of them being $u_3=0$~--,
since $(\vec a^2)^2-4 \vec b^2+4u_0^2\geq 0$.
The situation is similar to
the one-angular momentum case discussed in \cite{russo,csaki,minahan}.
The case $a_1=a_2 $, $u_0=0$ is special. This gives a double root with
$u_1=u_2=a_1^2$, so that $s_1=2$, $r_1=3$, $s_2=r_2=0$ and $s_2-s_1+2= 0$
and
hence the WKB method breaks down (cf. eq. \eqn{asss}).
breaks down. The WKB method breaks down also in the case
$a_1=a_2=a_3\gg u_0$ (see also below).

The formula for the mass spectrum (\ref{qcd3xi}) for $0^{++}$
glueballs and  its resonances implies an  important prediction
for the ratio between masses of two arbitrary resonances:
it is independent of $a_i$ (up to corrections of order $1/m^2$),
depending only on the radial quantum numbers
\be
{M^2_{m}\over M^2_{m'} } \cong {m(m+1)\over m'(m'+1)} \ .
\label{coomp}
\ee
This was observed in \cite{minahan}
for the case of ${\rm QCD}_4$ with one angular-momentum.
It implies that glueball masses with $m\gg 1$
vary only slightly in the whole
parameter space $a_1,a_2,a_3$.

%We also note that it is straightforward
%to extend our analysis to $0^{--}$ glueballs
%by solving the supergravity equations corresponding to 2-forms.

%%%%%%%%%%%%%%%%%%%%%%%%%%%%%%%%%%%%%%%%%%%%%%%%%%%%%%%%%%%
%%%%%%%%%%%%%%%%%%%%%%%%%%%%%%%%%%%%%%%%%%%%%%%%%%%%%%%%%%%
\subsection{Kaluza--Klein modes with $\tau$ dependence}
%%%%%%%%%%         %%%%%%%%%%%%%%%%%%%%%%%         %%%%%%%%

For particles with $U(1)$ charge associated with the circle parametrized by
$\tau $ one can check that
there exist simple (angular-independent) solutions of the form
\be
\Psi= \phi(U) e^{ik\cdot x} e^{2\pi i n T_H \tau}\ .
\label{anztau}
\ee
Substituting this into eq.~(\ref{dill}) we obtain
\ba
&& \del_U U^5 \left(\prod_{i=1}^3 \Big(1-{a_i^2\ov U^2}\Big)
- {U_0^4\ov U^4}\right)\del_U \phi
+R^2\left(U M^2 R^2 -4\pi^2 n^2 T_H^2 U^3 \D^{1/2} G^{\tau\tau} \right) \phi
=0  ,
\label{eqqtau}
\ea
where $G^{\tau\tau}$ is given in eq.~(\ref{giin}). Introducing a new radial
coordinate $u=U^2$ we obtain eq.
(\ref{diiff}) with $M^2$ replaced by $M^2- 4 \pi^2 n^2 T_H^2$ and
\ba
&& f = u^3-\vec a^2 u^2 +(\vec b^2 -u_0^2)u - \lqq \ , \qq h={R^4\ov 4}\ ,
\nonumber \\
&&p = - {R^4 \pi^2 n^2 T_H^2 u_0^2 u \ov
u^3 - \vec a^2 u^2 +(\vec b^2 -u_0^2)u - \lqq}\ .
\label{eqq2}
\ea

\no
Note that this equation is invariant under cyclic permutations of
$(a_1,a_2,a_3)$.
We find
\ba
&& s_1 =1\ ,\quad s_2=0\ ,\quad s_3 =-1\ ,\quad r_1 = 3\ , \quad
r_2 = 0 \ ,\quad r_3 = -2 \ ,
\nonumber \\
&& \a_1 =1\ ,\quad \a_2 = {2 R^2 \pi n T_H u_0 \sqrt{u_H}\ov
(u_H- u_1)(u_H-u_2)}\ ,
\quad \b_1 = 1\ ,\quad \b_2 =2 \ .
\label{fjk}
\ea

\no
Using \eqn{haw} we see that $\alpha_2=n$. Then \eqn{mxi}
(with $M^2\to M^2- 4 \pi^2 n^2 T_H^2$) gives
the formula
\be
M^2 = 4 \pi^2 n^2 T_H^2\ +\ {\pi^2\ov \xi^2 }\ m
(m+1+ n ) \ + \
{\cal O}(m^0)\ ,\qq m\ge 1\ ,
\label{tkkm}
\ee
where $\xi$ is given by \eqn{ashg}.
Using \eqn{jsdfg} we obtain that WKB is valid
%for the case of 1 real and 2 complex roots
when
\be
m(m+1+n) \gg (u_H-u_1)^{1/2} (u_H-u_2)^{1/2} {n^2 K^2(k)\ov u_H R^4 }
\quad {\rm and} \quad
{u_H K^2(k)\ov (u_H-u_1)^{1/2}(u_H-u_2)^{1/2} }\ .
\label{askc}
\ee
%
%Since the function $(1-k^2) K^2(k)$
%is bounded between $\pi^2/4$ and $0$

In order to obtain ${\rm QCD}_3$ through a
dimensional reduction of ${\rm QCD}_4$,
it is necessary that the mass scale for these Kaluza--Klein states be much
larger than that for the glueball masses.
{}From eq.~(\ref{tkkm}) one sees that this
requires the condition
\be
T_H\gg {1\over\xi }\ .
\ee
Using \eqn{haw} and \eqn{ashg}
this becomes
\be
K(k){u_H\over u_0} \Big[ \big(1-{u_1\over u_H}\big)
\big(1-{u_2\over u_H}\big)  \Big] ^{3/4} \gg 1\ ,
\label{decco1}
\ee
where $k$ is given by \eqn{ashg}.
$K(k)$ cannot be too large, otherwise the
WKB approximation breaks down (see \eqn{askc}).
Therefore eq.~(\ref{decco1}) implies that in order to decouple
the $\tau$ Kaluza--Klein
modes one needs that $u_H\gg u_0$. With no  loss of generality
we can assume that $a_1\geq a_2\geq a_3$.
{}It is clear from the equation for the horizon  that
one has $u_H\gg u_0$ if and only if  $u_H\simeq a_1^2 $.
Thus the region where these Kaluza--Klein particles decouple is
$a_1^2\gg u_0$.
To ensure at the same time  the validity
of the WKB approximation  \eqn{askc}
one needs that $u_H$ be not too close to $u_1$, which for
$u_H\gg u_0$ amounts to saying
that $a_1^2$ should not be too close to $a_2^2$.

%%%%%%%%%%%%%%%%%%%%%%%%%%%%
\subsection{Kaluza--Klein modes of $S^5$}
%%%%%%%%%%%%%%%%%%%%%%%%%%%%

The isometry group of the metric with $a_1=a_2=a_3=0$ contains
a factor $SO(6)$.
We will investigate the problem of decoupling for the $l=1$ Kaluza--Klein
modes
--~corresponding to the $\bf 6$ representation of $SO(6)$~--,
as modes with larger value for $l$ are expected to be heavier.
The three angular momenta break the symmetry down to the Cartan subgroup
$SO(2) \times SO(2) \times SO(2)$.
With respect to the Cartan subgroup, the representation ${\bf 6}$ decomposes
into
three doublets of $SO(2)$, i.e.
${\bf 6}\to ({\bf 2},{\bf 1},{\bf 1}) \oplus ({\bf 1},{\bf 2},{\bf 1})\oplus
({\bf 1},{\bf 1},{\bf 2})$.
These give rise to three equations,
which must be related by cyclic permutations of $(a_1,a_2,a_3)$.
For the first doublet, we make the ansatz
\be
\Psi = \phi(U) e^{ik\cdot x} \sin\th
\pmatrix{\cos\varphi_1\cr \sin\varphi_1}\ ,
\label{anz1}
\ee
and insert it into the Laplace equation (\ref{dill}).
One then obtains for $\phi(U)$ the differential equation
\ba
&& \del_U U^5 \left(\prod_{i=1}^3 \Big(1-{a_i^2\ov U^2}\Big)
- {U_0^4\ov U^4}\right)\del_U \phi + M^2 R^4 U \phi
\nonumber \\
&& + \Biggl( U^3\left(\D_2(\cot^2\th-4) - R^2 \D^{1/2}
G^{\varphi_1\varphi_1}\right)- 2 (a_2^2-a_3^2) U \cos 2\psi \Biggr)\phi= 0\
,
\label{eeqq1}
\ea
where $M^2=-k^2$.
After changing variable $u=U^2$, and a somewhat lengthy computation, we see
that all
$\th$ and $\psi$ dependence cancels out, and we obtain an equation of the
form
(\ref{diiff}) with
\ba
&&f= u^3-\vec a^2 u^2 +(\vec b^2 -u_0^2)u - \lqq \ ,\qq h={R^4\ov 4}\ ,
\nonumber \\
&&p = {1\ov 4 }\ {-5 u^4 + c_3 u^3 + c_2 u^2 + c_1 u + c_0
\ov u^3-\vec a^2 u^2 +(\vec b^2 -u_0^2)u - \lqq }\ ,
\label{eqq1}
\ea
where
\ba
c_3 & =& 4(2 \vec a^2-a_1^2)\ =\ 4(a_1^2+ 2 a_2^2 +2 a_3^2) \ ,
\nonumber \\
c_2 & = & -3 \left( (\vec a^2)^2 - a_1^4 + 2 b_1^2\right) + 5 U_0^4 \ =\
-6 a_1^2 a_2^2 - 3 a_2^4 - 6 a_1^2 a_3^2 -12 a_2^2 a_3^2 - 3 a_3^4 +
5 u_0^2\ ,
\nonumber \\
c_1 &=& 8 \lqq + (4 b_1^2- 3 u_0^2)(\vec a^2 - a_1^2) + 2 a_1^2 \left(
(\vec a^2)^2 - 2 \vec b^2 - a_1^4 \right)
\nonumber \\
&= & 2 a_1^2 a_2^4 + 8 a_1^2 a_2^2 a_3^2 + 4 a_2^4 a_3^2 + 2 a_1^2 a_3^4
+4 a_2^2 a_3^4 -3 (a_2^2 +a_3^2) u_0^2 \ ,
\nonumber\\
c_0 & = & -b_1^2 (2 \vec b^2 - b_1^2 -u_0^2)\ =\
- a_2^2 a_3^2 (2 a_1^2 a_2^2 + 2 a_1^2 a_3^2 + a_2^2 a_3^2 - u_0^2) \ .
\label{deqq1}
\ea

\no
When $a_2=a_3=0$, the differential equation reduces to eq.~(3.15)
of \cite{csaki} (with $u\to u^2$).
For $a_1=a_3=0$, it reduces to eq.~(3.16) of \cite{csaki}, which corresponds
to the other doublet (this is because  interchanging doublets is equivalent
to permuting $a_1,a_2,a_3$).
For the various constants of Sect.~3 we find
\ba
&& s_1=1 \ ,\quad s_2 =0 \ , \quad s_3 =- 1\ , \quad r_1 =3 \ ,\quad
r_2 =0 \ ,\quad r_3 =1 \ ,
\nonumber \\
&& f_1=(u_H-u_1)(u_H-u_2) \ ,~~ f_2 =1 \ , ~~ p_1 =
-{a_1^2 (u_H-a_2^2)^2 (u_H-a_3^2)^2 \ov 4 u_H f_1  } \ ,
~~ p_2=-{5\ov 4}\ ,
\nonumber \\
&& \a_1=1 \ ,\quad \a_2 = {a_1 (u_H -a_2^2)(u_H -a_3^2)\ov \sqrt{u_H}
f_1  }\ ,
\quad \b_1=1\ , \quad \b_2 =3 \ .
\label{sajh}
\ea

\no
Hence, using \eqn{jsdfg}, we find the mass
\be
M^2= {\pi^2\ov \xi^2}\ m\left(m+2 +
{a_1 (u_H -a_2^2)(u_H -a_3^2)\ov \sqrt{u_H} (u_H-u_1)(u_H-u_2)  } \right) \
+\
{\cal O}(m^0)\ ,\qq m\ge 1\ ,
\label{hsj}
\ee
where $\xi $ is given by \eqn{ashg}.
%In the case of two complex roots
%$(u_H- u_2)(u_H - u_3)$ should be replaced
%by $|u_H -u_1|^2$ in the above expressions (with $\xi$
%given by \eqn{ashg}).

Let us examine  if these states may have a large mass in the same
region $a_1^2 \gg u_0$
where the Kaluza--Klein modes with $\tau $ dependence decouple.
In this limit one has
$u_H\cong a_1^2$, $u_1\cong a_2^2, \ u_2\cong a_3^2$, and
the mass formula \eqn{hsj} takes the form
\be
M^2\simeq {\pi^2\ov \xi^2}\ m (m+ 3)\ + \ {\cal O}(m^0)\ ,\qq m\ge 1\ ,
\label{sqs}
\ee
where $\xi$ is again given by \eqn{ashg}.
This shows that for $a_1^2\gg u_0$ the mass of these Kaluza--Klein
states is of the same order as the glueball masses \eqn{qcd3xi}.
More generally, one can show that in every region
of the parameter space the Kaluza--Klein masses \eqn{hsj}
are of order $M= {\cal O}(1/\xi )$.
Indeed, it could only be otherwise in a region where $u_H\cong u_1$,
where the third term of  \eqn{hsj}   has a potential singularity,
but this happens only in the region $a_1^2\gg u_0$, 
which leads to \eqn{sqs}.

For the ansatz
\be
\Psi= \phi(U) e^{ik\cdot x} \cos\th \sin\psi
\pmatrix{\cos\varphi_2\cr \sin\varphi_2}\ ,
\label{anz2}
\ee
and for
\be
\Psi= \phi(U) e^{ik\cdot x} \cos\th \cos\psi
\pmatrix{\cos\varphi_3\cr \sin\varphi_3}\ ,
\label{anz3}
\ee
corresponding to the other two doublets,
we obtain the cyclic permutation in $(a_1,a_2,a_3)$ of (\ref{eqq1}),
as expected.
Note that the constants $c_0,\dots , c_3$ in
(\ref{deqq1}) are not invariant under such permutations.
The  constants of \eqn{sajh}
are obtained  by cyclic permutation
in $(a_1,a_2,a_3)$ of \eqn{sajh}.
For completeness we include the corresponding mass spectra. For the
Kaluza--Klein doublet \eqn{anz2} it is given by
\be
M^2= {\pi^2\ov \xi^2}\ m\left(m+2 +
{a_2 (u_H -a_3^2)(u_H -a_1^2)\ov \sqrt{u_H} (u_H-u_1)(u_H-u_2)  } \right) \
+\
{\cal O}(m^0)\ ,\qq m\ge 1\ ,
\label{hsj1}
\ee

\no
and for the Kaluza--Klein doublet \eqn{anz3} it is given by
\be
M^2= {\pi^2\ov \xi^2}\ m\left(m+2 +
{a_3 (u_H -a_1^2)(u_H -a_2^2)\ov \sqrt{u_H} (u_H-u_1)(u_H-u_2)  } \right) \
+\
{\cal O}(m^0)\ ,\qq m\ge 1\ ,
\label{hsj2}
\ee

\no
where $\xi $ is given by \eqn{ashg} in both mass formulae.
In the region $a_1^2\gg u_0$ where the circle Kaluza--Klein states decouple,
 eqs.~\eqn{hsj1} and \eqn{hsj2} take the
form
\be
M^2\simeq {\pi^2\ov \xi^2}\ m (m+ 2)\ + \ {\cal O}(m^0)\ ,\qq m\ge 1\ .
\label{sqs1}
\ee
These are essentially the same mass formulae as in the case of the
$0^{++}$ glueballs,
so that the corresponding mass scales are the same.

Finally, we note that the conditions for the
validity of the WKB approximation in the case of
the $S^5$ Kaluza--Klein modes are roughly the same as the corresponding ones
for the $0^{++}$ glueballs we have already discussed.
In all cases the WKB approximation can be applied everywhere, except in the
region $a_1^2=a_2^2\gg u_0$.

%%%%%%%%%%%%%%%%%%%%%%%%%%%%%%%%%%%%%

\section{Conclusions }

In this paper we have considered
non-supersymmetric QCD models in 2+1 dimensions based on an asymptotically
$AdS_5\times S^5$ static geometry constructed
from the rotating D3-brane with maximum number of rotation parameters.
Within the WKB approximation, closed
analytic formulas have been obtained
for the mass spectra of  $0^{++}$ scalar glueballs and
of states corresponding to excitations in the internal parts (circle and
sphere) of
the space.
The various mass spectra were found to depend, for every model and
to the first two leading orders in the WKB approximation, in a
universal manner on the rotation parameters:
modulo slight fluctuations
(such as the one produced by the third term in \eqn{hsj}),
only two mass scales appear, denoted $1/ \xi$ and $T_H$,
the latter characterizing the masses of Kaluza--Klein particles
associated with the $\tau $ direction,
the former dictating the masses of all other
scalar modes with vanishing charge in the $\tau $ direction.
This is a bit surprising, and it implies that,
 despite the large number of parameters,
the sphere Kaluza--Klein modes
--~unlike the ones corresponding to the circle~--
do not decouple in any region of parameter space.
%At most we found them to have slightly heavier masses
%than the corresponding scalar glueballs (for ${\rm QCD}_3$,
%compare \eqn{qcd3xi} with \eqn{sqs} and \eqn{sqs1} and
%for ${\rm QCD}_4$, compare \eqn{sadj} with
%\eqn{sqs3}, \eqn{sqs4} and \eqn{sq23}).
The sphere Kaluza--Klein states have no analogue at the weakly
coupled finite-$N$
pure $SU(N)$ Yang--Mills theories.
Masses computed in the supergravity approximation
 can in principle get important corrections upon extrapolating from the
strong-coupling  regime $\lambda\gg 1$ to the weak-coupling regime
$\lambda \ll 1$.
Comparison with lattice results \cite{COOT98,csaki} suggests
that singlet particles should only be slightly changed
in the extrapolation process, whereas non-singlet
particles should get large corrections.

The extremal solution with $u_0=0$ saturates the Bogomol'nyi bound
and has unbroken  supersymmetries.
The near-supersymmetric case $a^2_i\gg u_0$ is precisely
the interesting case where the radius of the extra brane direction
shrinks to zero, so that the associated Kaluza--Klein particles decouple
and  the system becomes effectively ($2+1$)-dimensional.
It would be very interesting to establish what are the
states whose masses  are protected by supersymmetry
in the rotating system with
$u_0=0$. This might explain
why glueball masses have values close to the values
obtained by lattice calculations \cite{csaki,teper}.

%%%%%%%%%%%%%%%%%%%%%%%%%%%%%%%%%%%%%
\bigskip\bigskip

\centerline{\bf Acknowledgements }

We have benefited from discussions with C.~Cs\'aki and J. Terning on
closely related matters. J.R. would like to thank
Fundaci\' on Antorchas for financial support (project A-13681/1).

\vfill\eject
%%%%%%%%%%%%%%%%%%%%%%%%
% ---- Bibliography ----

\end{document}